\documentclass[twocolumn]{aastex631}

\newcommand{\src}{S4 0954+65}

\usepackage[dvipsnames]{xcolor}
\definecolor{MyBlue}{RGB}{0,0,255}
\usepackage{hyperref}
\usepackage{subfigure}
\usepackage{longtable}
\usepackage{makecell}
\usepackage{tabularx}
\usepackage{amsmath}
\usepackage{enumitem}
\setlist{nosep}
\usepackage{booktabs}

\begin{document}

\title{Blazar \src\ in Outbursts: A Broadband View of High-Energy Flaring}

\author[0000-0001-9092-2903]{P. K. Navaneeth}
\affiliation{Doctoral School of Exact and Technical Sciences, University of Zielona Góra, 65-516 Zielona Góra, Poland; email: \textcolor{blue}{940d@g.elearn.uz.zgora.pl}}

\author[0000-0002-0705-6619]{Gopal Bhatta}
\affiliation{Janusz Gil Institute of Astronomy, University of Zielona Góra, 65-516 Zielona Góra, Poland; email: \textcolor{blue}{g.bhatta@ia.uz.zgora.pl}}



\begin{abstract}
We present a multi-wavelength study of the blazar {\src} from 2008 to 2026, focused on $\gamma$-rays and five high-energy flares. All flares show significant intra-day $\gamma$-ray variability, with shortest flux doubling/halving timescales of 3.5--7.8\,hr. The fractional variability increases with the energy of the radiating electrons, from a minimum in X-rays to $F_{\rm var}=(106.5\pm1.1)\%$ in $\gamma$-rays, as in flat-spectrum radio quasars (FSRQs). The lognormal flux distribution detected in all bands arises from intrinsic fluctuations in a particle-dominated emission region, combined with changes in its Doppler factor. Four photons of 123--188 GeV were detected in Fermi-LAT in intermediate and low states, indicating that emission above 100 GeV is not confined to outbursts and arises beyond the broad-line region. The X-ray observations show a weak harder-when-brighter trend, as in FSRQs. One-zone leptonic models of the flare spectral energy distributions attribute the $\gamma$-rays to external Compton scattering of dusty torus photons in compact, particle-dominated regions of $R=(0.71$--$1.4)\times10^{16}$ cm with bulk Lorentz factors of 15.3--28.6, located $(0.34$--$4.4)\times10^{18}$ cm from the black hole, close to the torus. The external Compton origin of the $\gamma$-rays is typical of FSRQs, whereas the synchrotron peak and Compton dominance are intermediate between the two blazar classes. The {\src} is therefore a transitional blazar with an FSRQ-like central engine.
\end{abstract}



\section{Introduction} \label{sec:intro}
Blazars are a subclass of radio-loud active galactic nuclei (AGN) characterised by highly relativistic jets oriented at small angles to the observer's line of sight \citep{Urry1995, Blandford1977}. Their broadband emission, spanning from radio to very-high-energy (VHE, $E > 100\,\mathrm{GeV}$) $\gamma$-rays, is amplified by relativistic Doppler boosting \citep{Blandford1979}, making blazars among the most luminous persistent sources in the universe. They display extreme observational characteristics, including rapid flux variability on timescales from minutes to years \citep{Heidt1996, Bhatta2021Dec}, high optical polarisation, superluminal motion, and intense $\gamma$-ray emission \citep{Miller1978, Abdo2010}.

The broadband spectral energy distribution (SED) of blazars exhibits a characteristic double-humped structure. The low-energy hump, peaking between radio and X-ray frequencies, is attributed to synchrotron radiation from relativistic electrons in the jet \citep{Urry1998}. The origin of the high-energy hump, extending into the GeV--TeV range, remains a subject of active research. In leptonic models, this emission arises from inverse Compton (IC) scattering of low-energy photons, with seed photons originating either from within the jet in the synchrotron self-Compton (SSC) scenario \citep{Maraschi1992, Marscher1985}, or from external fields such as the accretion disk, broad-line region (BLR), or dusty torus in the external Compton (EC) scenario \citep{Dermer1992, Sikora1994, Ghisellini1996}. Hadronic models attribute the high-energy emission to proton synchrotron radiation or secondary particles from photo-pion and photo-pair interactions \citep{Aharonian2002, Bottcher2013}.

Blazars are broadly classified by two schemes. Based on optical spectral properties, they are divided into Flat Spectrum Radio Quasars (FSRQs), which display prominent broad emission lines with rest-frame equivalent width $\mathrm{EW} > 5\,\text{\AA}$, and BL Lacertae objects (BL Lacs), which exhibit featureless or weak-line spectra with $\mathrm{EW} < 5\,\text{\AA}$ \citep{Stocke1991, Urry1995}. This distinction broadly corresponds to different accretion regimes, with FSRQs associated with radiatively efficient accretion and BL Lacs with radiatively inefficient modes \citep{Ghisellini2011Jul}. Blazars are further categorised by the frequency of their synchrotron peak ($\nu_{\mathrm{p}}$) into Low Synchrotron Peaked (LSP; $\nu_{\mathrm{p}} \leq 10^{14}$ Hz), Intermediate Synchrotron Peaked (ISP; $10^{14} < \nu_{\mathrm{p}} < 10^{15}$ Hz), and High Synchrotron Peaked (HSP; $\nu_{\mathrm{p}} \geq 10^{15}$ Hz) sources \citep{Abdo2010}.

{\src} (RA: 149.6968, Dec: 65.5652) is a blazar located at a redshift of $z = 0.3694 \pm 0.0011$ \citep{Lawrence1986Mar, BecerraGonzalez2021Jul}. Originally detected as a radio source in the Jodrell-Bank 996 MHz Survey, its optical counterpart identified by \citet{Cohen1977Jan} and was classified as a BL Lac object by \citet{Walsh1984Nov} based on its featureless optical continuum. It was subsequently identified as an extragalactic $\gamma$-ray emitter by EGRET, through its optical and radio association \citep{Mukherjee1995May}. The classification of this source has been debated in the literature. It is listed as a BL Lac object in the Fermi-LAT 4LAC \citep{Ajello2020Apr} and CGRaBS \citep{Healey0803} catalogs, yet several of its emission properties are typical of FSRQs. The optical spectrum is state dependent with the broad Mg\,{\sc ii} line and the narrow O\,{\sc iii} and O\,{\sc ii} lines are detected only in low states, while in high states the jet continuum dominates and the lines fall below the detection limit \citep{BecerraGonzalez2021Jul}. In addition, the broadband SED shows a high Compton dominance that standard one-zone SSC models cannot accommodate, which requires EC scattering as is typical of FSRQs \citep{Tanaka1608, MAGICCollaboration2018Sep}. This ambiguity has led to the source being regarded as a transitional blazar with characteristics intermediate between the BL Lac and FSRQ classes \citep{Tanaka1608, Ghisellini2011Jul, BecerraGonzalez2021Jul}. The source is well known for its high variability. Intra-day variability has been reported in the optical and radio bands since the early 1990s, particularly during flaring states \citep{Wagner1990Aug, Wagner1993Apr, Raiteri1999Dec, Raiteri2021Jul}. In 2022, it reached its historical optical maximum \citep{Raiteri2023Dec}. The optical polarisation is also highly variable \citep{Kuehr1990Jan, Stickel1991Jun}, and large rotations of the polarisation angle have been accompanied by the emergence of new superluminal knots in the parsec-scale jet \citep{Morozova2014Sep, Morozova2016Sep, MAGICCollaboration2018Sep}. Based on the location of its synchrotron peak, the source is classified as an LSP in the 4LAC catalog \citep{Ajello2020Apr}. In February 2015, an exceptionally high optical and near-infrared flaring state triggered follow-up observations by the MAGIC telescopes, leading to the discovery of VHE $\gamma$-ray emission at a significance of $\sim7.4\sigma$ \citep{Mirzoyan2015ATel, MAGICCollaboration2018Sep}.

In this work, we study the multi-wavelength nature of the {\src} and its high energy flares through timing and spectral analysis. This paper is organised as follows: Section \ref{sec:data} describes the multi-wavelength observations and data reduction procedures. Section \ref{sec:analysis} presents the long-term variability analysis, with a focus on the $\gamma$-rays, followed by the spectral analysis of the high-energy flares and the broadband SED modelling. Section \ref{sec:discussion} discusses the physical interpretation of our results, and Section \ref{sec:conclusion} summarises the main conclusions.

\section{Data Reduction} \label{sec:data}

\subsection{Fermi-LAT}
Fermi-LAT data were obtained from the Fermi Science Support Center\footnote{\url{https://fermi.gsfc.nasa.gov/ssc/data/access/lat/}} for a 20$^{\circ}$ region of interest (ROI) centred on the source position. The analysis used Pass 8 (P8R3) data \citep{Atwood2013} in the 0.1--300 GeV energy range, with FermiTools 2.2.0 and Fermipy 1.2.2 \citep{Wood2017}. SOURCE class events (evclass = 128, evtype = 3) were selected within a $10^{\circ}$ ROI using the P8R3\_SOURCE\_V3 instrument response functions. Standard data quality cuts were applied, including a zenith angle cut of $z_{\rm max} < 90^{\circ}$ to minimise Earth-limb contamination, and good time intervals were filtered using \texttt{(DATA\_QUAL>0)\&\&(LAT\_CONFIG==1)}. The $\gamma$-ray signal significance was quantified via the maximum-likelihood test statistic TS $= 2(\log\mathcal{L}_1 - \log\mathcal{L}_0)$, where $\mathcal{L}_1$ and $\mathcal{L}_0$ are the likelihoods with and without a point source at the target position \citep{Mattox1996}.

The source model included all 4FGL catalog sources \citep{Abdollahi2020} within $15^{\circ}$ of the ROI center, together with the Galactic (\texttt{gll\_iem\_v07}) and isotropic diffuse (\texttt{iso\_P8R3\_SOURCE\_V3\_v1}) components. A binned likelihood analysis was performed with eight energy bins per decade and $0.1^{\circ}$ pixel binning. Following initial optimisation, sources with TS $<$ 10 were fixed, while the normalisations of sources with TS $>$ 10 and all sources within $3^{\circ}$ of the ROI centre were left free, along with the diffuse components. A weekly binned light curve was produced, retaining only time bins with TS $>$ 9 as statistically significant detections. 

\begin{table}
\caption{The high-energy flaring epochs of {\src} identified from the $\gamma$-ray and X-ray light curves.\label{tab:flares}}
\centering
\let\tabular\savetabular
\begin{tabular}{lcc}
    \toprule
    Epoch & Time period & Duration \\
     & (MJD) & (days) \\
    \midrule
    F1 & 57048--57090 & 42 \\
    F2 & 59680--59715 & 35 \\
    F3 & 59717--59746 & 29 \\
    F4 & 60086--60103 & 17 \\
    F5 & 60695--60730 & 35 \\
    \bottomrule
\end{tabular}
\end{table}

\subsection{Swift XRT \& UVOT} \label{subsec:uvot}
The X-ray spectra and light curve of the source were generated using the Swift-XRT Online Data Products Generator\footnote{\url{https://www.swift.ac.uk/user_objects/}}, following the data reduction procedures described in \cite{Evans2007, Evans2009}. The 0.3-10 keV spectra were grouped to a minimum of one count per energy bin and fitted with different spectral models in XSPEC\footnote{\url{https://heasarc.gsfc.nasa.gov/xanadu/xspec/}} \citep{Arnaud1996} version 12.14.0 using Cash statistics \citep{Cash1979}. The Galactic hydrogen column density ($n_{\rm H}$) was fixed to $4.68 \times 10^{20}\ \rm{cm}^{-2}$ \citep{Willingale2013}. 

\begin{figure*}[!t]
    \centering
    \includegraphics[width=1\linewidth]{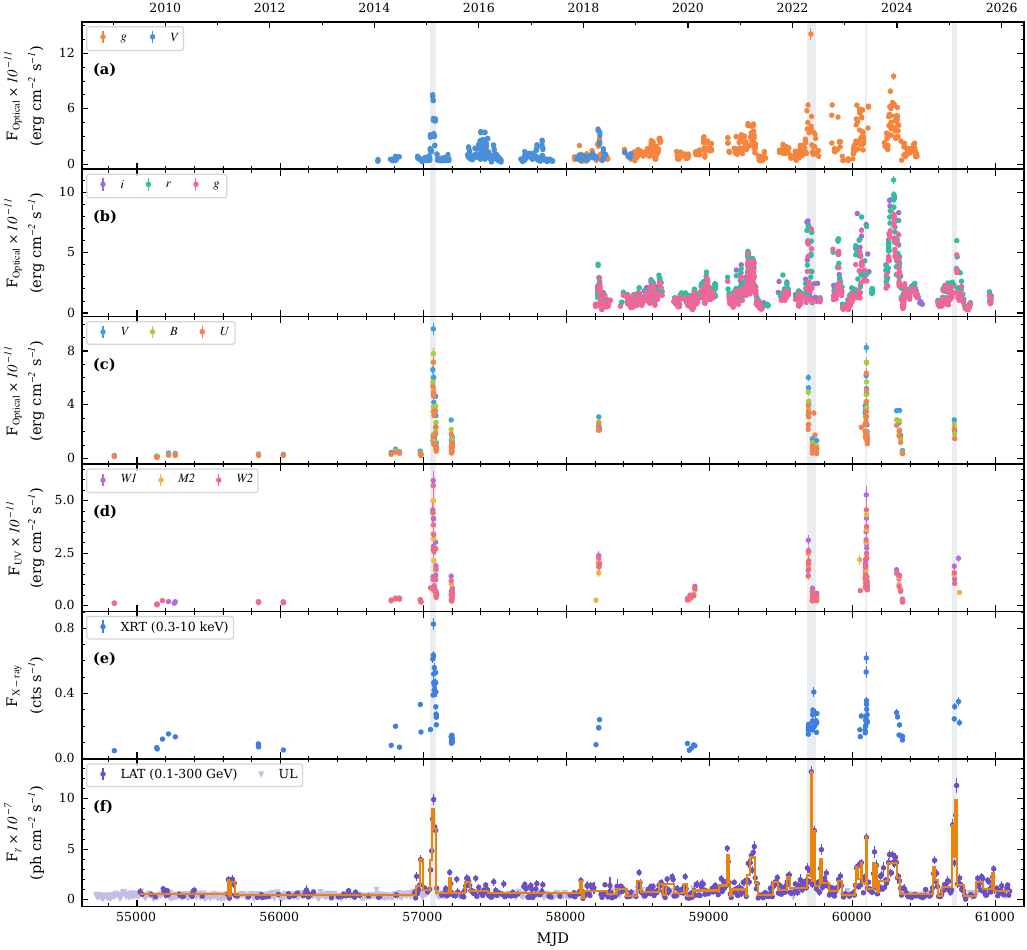}
    \caption{Multi-wavelength light curve of the blazar {\src}: (a) ASAS-SN optical $g$ and $V$ bands; (b) ZTF optical $i$, $r$, and $g$ bands; (c) Swift-UVOT optical $V$, $B$, and $U$ bands; (d) Swift-UVOT ultraviolet $W1$, $M2$, and $W2$ bands; (e) Swift-XRT X-ray in the 0.3–10 keV band; (f) Fermi-LAT $\gamma$-ray in the 0.1–300 GeV band, with the orange line indicating the Bayesian-block segmentation. The grey shaded region represents the five flaring epochs selected in this work.}
    \label{fig:mwl_lc}
\end{figure*}

The source was observed with six standard UVOT photometric filters (optical: B, V, U; UV: UVM2, UVW1, UVW2). The observations were downloaded from the SWIFTMASTER catalogue, and data analysis was performed using HEASoft version 6.33 and CALDB version 20220331, following the standard procedures of the Swift UVOT Software Guide\footnote{\url{https://www.swift.ac.uk/analysis/uvot}}. Source counts were extracted from a $6^{\prime\prime}$ radius aperture centred on the source, while background counts were measured from a $20^{\prime\prime}$ radius circle in a nearby source-free region. Images across all extensions were combined for each observation using \texttt{uvotimsum}, and aperture photometry was performed using \texttt{uvotsource} for each filter independently. Galactic extinction corrections were applied using $E(B-V) = 0.1015 \pm 0.0118$ from \cite{Schlafly2011} and the extinction laws of \cite{Cardelli1989}, and the corrected magnitudes were converted to flux densities using zero points from \cite{Breeveld2011} and conversion factors from \cite{Poole2007}.

\subsection{Archival Data}
Long-term optical coverage was supplemented with archival photometry from the All-Sky Automated Survey for Supernovae (ASAS-SN) \citep{Shappee2014, Hart2023} and the Zwicky Transient Facility (ZTF) \citep{Masci2018, Bellm2018}. ASAS-SN light curves were retrieved from the Sky Patrol V2.0 photometry database\footnote{\url{http://asas-sn.ifa.hawaii.edu/skypatrol}}, in the $V$ band up to late 2018 and in the $g$ band thereafter. ZTF $g$, $r$ and $i$ band light curves derived from point spread function (PSF) fitting photometry were obtained from ZTF Data Release\footnote{\url{https://www.ztf.caltech.edu/}}. Both data sets were filtered using the instrument-specific quality flags and corrected for Galactic extinction as described in Section \ref{subsec:uvot}, and converted to flux densities.

\section{Results \& Analysis} \label{sec:analysis}

\subsection{Multi-Wavelength Light Curves and \\ High-Energy Flares} \label{sec:mwllc}
The source exhibits variability in all wavebands, with multiple flares over the observation period as shown in Figure \ref{fig:mwl_lc}. The highest $\gamma$-ray flux in the weekly light curve is $1.27\times10^{-6}$\,ph\,cm$^{-2}$\,s$^{-1}$, on 2022 May 12 (MJD 59711.5) and the highest X-ray count rate measured is 0.83\,cts\,s$^{-1}$, on 2015 February 17 (MJD 57070.8). In the optical and ultraviolet bands, the highest fluxes are $1.41\times10^{-10}$\,erg\,cm$^{-2}$\,s$^{-1}$ on 2022 May 9 (MJD 59708.2) measured in the ASAS-SN $g$ band, and $5.97\times10^{-11}$\,erg\,cm$^{-2}$\,s$^{-1}$ on 2015 February 16 (MJD 57069.8) measured in the UVOT $W1$ band. 

To identify the flaring epochs of the source, the Bayesian Blocks (BB) algorithm \citep{Scargle1302} was employed together with the group-finding algorithm HOP \citep{Eisenstein9805}, following its application to light curves by \citet{Meyer2019May} and \citet{Wagner2203}. The BB algorithm provides a blockwise constant representation of a light curve by identifying statistically significant variations, with each block characterised by average flux. We adopted the Astropy \citep{AstropyCollaboration2013Oct, AstropyCollaboration2018Sep, AstropyCollaboration2022Aug} implementation with a false-alarm probability of $p_0 = 0.05$. The HOP algorithm was then used to define the duration of the flares and to separate overlapping ones. For the $\gamma$-ray band, the BB and HOP analyses were applied to the weekly binned light curve. To select the major outbursts, only the HOP groups with mean flux $F_{\mathrm{HOP}} \geq 5 \times \bar{F}$ were chosen, where $\bar{F}$ is the mean flux of the light curve. For the X-ray band, the flares were selected directly from the BB representation of the light curve, as the irregular and sparse sampling does not provide meaningful HOP grouping. We focused on the high-energy flares with simultaneous multi-wavelength data: three identified in the $\gamma$-ray band, F1 (2015 January 26 to March 9), F2 (2022 April 11 to May 16) and F5 (2025 January 20 to February 24), and two identified in the X-ray band, F3 (2022 May 18 to June 16) and F4 (2023 May 22 to June 8). The flaring epochs are shaded in the multi-wavelength light curves shown in Figure \ref{fig:mwl_lc} and listed in Table \ref{tab:flares}. The $\gamma$-ray and optical maxima both fall within the F2 epoch, while the X-ray and ultraviolet maxima fall within F1 together with the highest UVOT flux $9.64\times10^{-11}$\,erg\,cm$^{-2}$\,s$^{-1}$ in $V$ band.

\subsection{Variability Timescale} \label{sec:tvar}
To quantify the rapid flux variability of the source in $\gamma$-rays, the shortest flux doubling (or halving) timescale is computed assuming an exponential evolution of the flux: 
\begin{equation}
    F(t) = F(t_0)\, 2^{-(t - t_0)/\tau},
    \label{eq:tvar}
\end{equation}
where $F(t)$ and $F(t_0)$ are the fluxes at times $t$ and $t_0$, respectively, and $\tau$ is the doubling/halving timescale \citep{Foschini1106, Prince2021b, Acharyya2306}. 

For each flare, a 12 hr binned light curve was generated, and the doubling/halving timescale was computed for all pairs of detections in the light curve. The results are summarised in Table \ref{tab:tvar}. Short-term variability is found in all five flares, with the timescales ranging from $\tau = 3.52 \pm 0.73$ hr in F2 to $\tau = 7.83 \pm 2.54$ hr in F1. In four of the five flares the variability corresponds to flux-halving (decay) with the exception is F3, where its flux-doubling (rise). These timescales indicate that the source exhibited statistically significant intra-day variability (IDV) in the GeV band during each of the flaring epochs. Hour-scale $\gamma$-ray variability has been reported for several FSRQs, including 3C 454.3, 3C 273 and PKS B1222+216 \citep{Foschini1106}, PKS 1510$-$089 \citep{Tavecchio2010Jun, Saito1303}, and 3C 279 and CTA 102 \citep{Meyer2019May}.

\begin{figure*}
    \centering
    \includegraphics[width=1.0\linewidth]{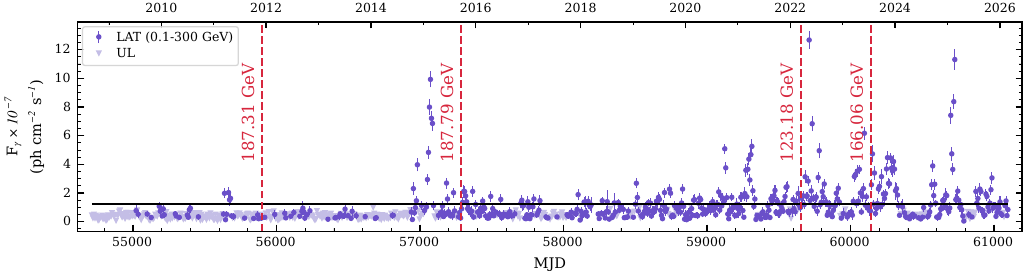}
    \caption{Fermi-LAT $\gamma$-ray light curve of the blazar {\src} with the arrival times of the highest-energy photons (HEPs) detected.}
    \label{fig:lat_hep}
\end{figure*}

The shortest variability timescale constrains the size of the emission region through light-crossing causality argument. For emission originating from a relativistic blob, the radius of the region satisfies $R \lesssim c\,\tau\,\delta_D/(1+z)$, where $c$ is the speed of light and $\delta_D$ is the Doppler factor. The timescales in Table \ref{tab:tvar} gives $R \lesssim (2.8-6.2)\times10^{15}\,(\delta/10)$ cm. $\delta_D \approx 9-11$ have been derived for the source from VLBA kinematics and radio variability \citep{Jorstad2017Sep, Liodakis2023Dec}, for which $R$ correspond to $\sim10^{-3}$ pc, smaller than the structures characterised in 43 GHz VLBA images \citep{Jorstad2017Sep}. The $\gamma$-ray emission originates in a compact region, whose $R$ and $\delta_D$ are further constrained by the SED modelling.

\begin{table}
\caption{Shortest flux doubling/halving timescales measured for {\src} in the 12~hr binned Fermi-LAT flare light curves.\label{tab:tvar}}
\centering
\let\tabular\savetabular
\begin{tabular}{lccc}
    \toprule
    Epoch & $\tau$ & Type & Time period \\
     & (hr) & & (MJD) \\
    \midrule
    F1 & $7.83 \pm 2.54$ & Decay & 57085.75--57086.75 \\
    F2 & $3.52 \pm 0.73$ & Decay & 59708.75--59709.25 \\
    F3 & $4.87 \pm 1.59$ & Rise  & 59729.25--59729.75 \\
    F4 & $7.50 \pm 1.77$ & Decay & 60095.75--60096.75 \\
    F5 & $5.06 \pm 1.51$ & Decay & 60701.75--60702.25 \\
    \bottomrule
\end{tabular}
\end{table}

\subsection{Highest $\gamma$-ray photons} \label{sec:heps}
To identify the highest-energy $\gamma$-ray photons from the source, we repeated the likelihood fit using the stricter \texttt{ULTRACLEAN} event class (\texttt{evclass}=512). Each event and its probability $P$, of having originated from the source is calculated using the \texttt{compute\_srcprob} routine of \texttt{fermipy}, a wrapper of \texttt{gtsrcprob}, which computes this probability from the relative contributions of the fitted model components at the position and energy of the event. Among the events with $P > 0.99$, the highest-energy photon has an energy of 187.79 GeV and arrived on MJD 57287.743. All such photons above 100 GeV are listed with their arrival times and probabilities in Table \ref{tab:heps}. The reconstructed direction of each of these photons was verified to lie within $0.1^{\circ}$ of the source position.

\begin{table}
\caption{Highest $\gamma$-ray photons (${>}100$~GeV) associated with {\src}, ordered by energy, with the probability of originating from the source ($P>0.99$) and arrival time.\label{tab:heps}}
\centering
\let\tabular\savetabular
\begin{tabular}{ccc}
    \toprule
    $E$ & $P$ & Arrival time \\
    (GeV) & & (MJD) \\
    \midrule
    187.79 & 0.994 & 57287.743 \\
    187.31 & 0.999 & 55901.383 \\
    166.06 & 0.998 & 60144.086 \\
    123.18 & 0.999 & 59655.195 \\
    \bottomrule
\end{tabular}
\end{table}

To characterise the source when each photon is emitted, the arrival time was compared with the weekly-binned $\gamma$-ray light curve as shown in Figure \ref{fig:lat_hep}, with the flux of the bin containing each photon expressed relative to the mean flux $\bar{F} = 1.26 \times 10^{-7}~\mathrm{ph\,cm^{-2}\,s^{-1}}$. None of the four photons arrived within the flare epochs F1--F5. The $166$ and $123$~GeV photons were recorded in weeks with fluxes of $2.0\times\bar{F}$ and $1.3\times\bar{F}$, and thus in intermediate states. The two most energetic photons arrived in fainter states, the $187.8$~GeV photon in a week at $0.7\times\bar{F}$, and the $187.3$ GeV photon in a week for which the light curve yields only an upper limit because of the low photon statistics. The arrival of the most energetic photons in intermediate and low states indicates that the emission above 100 GeV is not confined to the outbursts. Photons of these energies are rarely recorded from FSRQs, in which the dense radiation fields around the jet can limit the particle acceleration to VHE \citep{Paliya2025Sep}. The $\gamma$-rays above 10 GeV produced within the BLR are absorbed through $\gamma\gamma$ pair production \citep{Liu2006Dec, Pacciani2014Aug, Bottcher2016Apr}. The detection of VHE emission from blazars with broad emission lines has been used to place their $\gamma$-ray emission region beyond the BLR \citep{Aleksic2011Mar, Tanaka2013Nov, Abeysekara2015Dec, MAGICCollaboration2018Nov, MAGICCollaboration2021Mar, Paliya2025Sep}. Since the presence of BLR, traced by its broad Mg\,\textsc{ii} line \citep{BecerraGonzalez2021Jul}, the detection of these photons places the region that emitted them beyond the BLR. 

\subsection{Flux Distribution} \label{sec:fluxdist}
The flux distribution of a light curve can provide information on the process that drives the variability. A normal distribution is expected when the variability results from a linear stochastic process, whereas a lognormal distribution with a multiplicative process \citep{Uttley2005May}. Each distribution was modelled with a normal probability density function (PDF),
\begin{equation}
f_{\rm N}(F) = \frac{1}{\sqrt{2\pi}\,\sigma}\exp\left[-\frac{(F-\mu)^{2}}{2\sigma^{2}}\right],
\label{eq:normal}
\end{equation}
where $\mu$ and $\sigma$ are the mean and standard deviation of the flux, and with a lognormal PDF,
\begin{equation}
f_{\rm LN}(F) = \frac{1}{\sqrt{2\pi}\,s\,F}\exp\left[-\frac{(\ln F-m)^{2}}{2s^{2}}\right],
\label{eq:lognormal}
\end{equation}
where $m$ and $s$ are the mean and standard deviation of $\ln F$. The parameter uncertainties are estimated with the moving-block bootstrap \citep{Kunsch1989Sep} method, which resamples blocks of consecutive points and thereby preserves the correlation between neighbouring measurements. The block length of each light curve was selected automatically from the logarithm of the flux following \citet{Politis2004} and \citet{Patton2009Jul}. The uncertainty of each parameter is the standard deviation over 1000 resampled light curves, each refitted with both models.

\begin{table*}
\caption{Lognormal and normal distribution fit results for the flux distributions of {\src} in each band, obtained with maximum likelihood estimation.\label{tab:fluxdist}}
\centering
\let\tabular\savetabular
\begin{tabular}{lcccccc}
\toprule
 & \multicolumn{3}{c}{Lognormal fit} & \multicolumn{3}{c}{Normal fit} \\
\cmidrule(lr){2-4} \cmidrule(lr){5-7}
Band & $m$ & $s$ & LL & $\mu$ & $\sigma$ & LL \\
\midrule
Fermi-LAT & $-16.25\pm0.09$ & $0.82\pm0.04$ & $8066.30$ & $(1.26\pm0.13)\times10^{-7}$ & $(1.38\pm0.20)\times10^{-7}$ & $7720.11$ \\
Swift-XRT & $-1.64\pm0.15$ & $0.63\pm0.08$ & $72.00$ & $(2.36\pm0.37)\times10^{-1}$ & $(1.53\pm0.28)\times10^{-1}$ & $48.93$ \\
Swift-UVOT $W2$ & $-25.67\pm0.10$ & $1.00\pm0.06$ & $2303.64$ & $(1.13\pm0.11)\times10^{-11}$ & $(1.11\pm0.13)\times10^{-11}$ & $2261.09$ \\
Swift-UVOT $M2$ & $-25.70\pm0.11$ & $0.99\pm0.06$ & $2234.49$ & $(1.09\pm0.11)\times10^{-11}$ & $(1.06\pm0.12)\times10^{-11}$ & $2194.37$ \\
Swift-UVOT $W1$ & $-25.58\pm0.10$ & $1.03\pm0.06$ & $2340.43$ & $(1.26\pm0.12)\times10^{-11}$ & $(1.23\pm0.13)\times10^{-11}$ & $2298.78$ \\
Swift-UVOT $U$ & $-25.27\pm0.10$ & $1.00\pm0.06$ & $2169.76$ & $(1.65\pm0.15)\times10^{-11}$ & $(1.49\pm0.15)\times10^{-11}$ & $2139.73$ \\
Swift-UVOT $B$ & $-25.14\pm0.10$ & $1.00\pm0.06$ & $2087.56$ & $(1.87\pm0.17)\times10^{-11}$ & $(1.68\pm0.16)\times10^{-11}$ & $2058.61$ \\
Swift-UVOT $V$ & $-24.90\pm0.11$ & $0.97\pm0.06$ & $2021.79$ & $(2.32\pm0.22)\times10^{-11}$ & $(2.03\pm0.20)\times10^{-11}$ & $1995.21$ \\
ASAS-SN $g$ & $-24.92\pm0.12$ & $0.65\pm0.06$ & $13255.06$ & $(1.89\pm0.26)\times10^{-11}$ & $(1.46\pm0.23)\times10^{-11}$ & $13037.68$ \\
ASAS-SN $V$ & $-25.42\pm0.10$ & $0.65\pm0.06$ & $8625.64$ & $(1.16\pm0.16)\times10^{-11}$ & $(1.04\pm0.25)\times10^{-11}$ & $8427.00$ \\
ZTF $g$ & $-25.12\pm0.11$ & $0.70\pm0.07$ & $18624.21$ & $(1.59\pm0.23)\times10^{-11}$ & $(1.33\pm0.27)\times10^{-11}$ & $18284.90$ \\
ZTF $r$ & $-24.90\pm0.11$ & $0.68\pm0.06$ & $20497.31$ & $(1.96\pm0.26)\times10^{-11}$ & $(1.57\pm0.31)\times10^{-11}$ & $20148.50$ \\
ZTF $i$ & $-24.58\pm0.11$ & $0.65\pm0.06$ & $3797.76$ & $(2.64\pm0.34)\times10^{-11}$ & $(1.92\pm0.30)\times10^{-11}$ & $3744.34$ \\
\bottomrule
\end{tabular}
\par\vspace{4pt}
\noindent\parbox{\linewidth}{\textsc{Note:} For the lognormal fit, $m$ and $s$ are the mean and standard deviation of the logarithm of the flux, and for the normal fit, $\mu$ and $\sigma$ are given in the flux units of the corresponding band. The lognormal model is preferred over the normal model in all bands by both the AICc and the BIC.}
\end{table*}

\begin{figure*}
    \centering
    \includegraphics[width=1\linewidth]{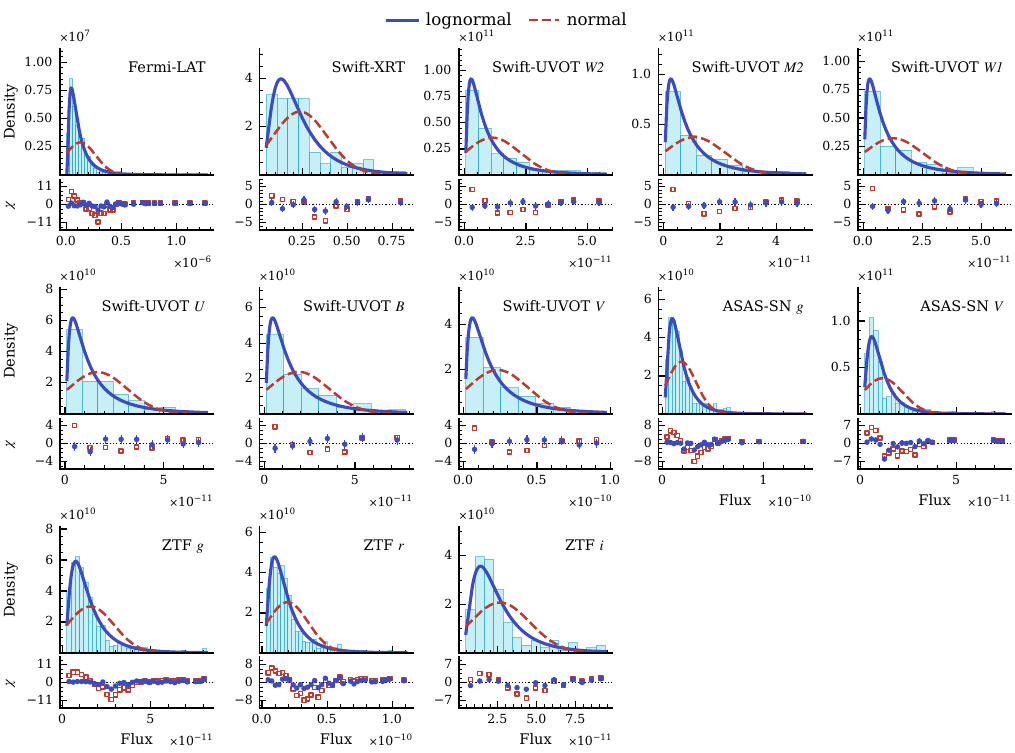}
    \caption{Flux distributions of the blazar {\src} in the multi-wavelength bands. In the upper sub-panel of each band, the histogram shows the probability density flux, and the solid blue and dashed red curves show the maximum-likelihood lognormal and normal fits, respectively. The lower sub-panel shows the residuals for the models in the same colours.}
    \label{fig:fluxdist}
\end{figure*}

\begin{figure}
    \centering
    \includegraphics[width=1\linewidth]{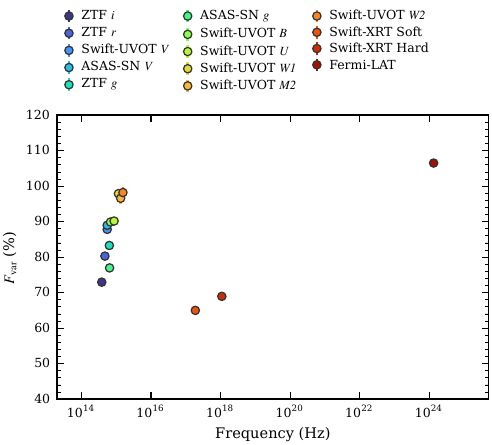}
    \caption{Fractional variability amplitude ($F_{\rm var}$) of {\src} across the optical, ultraviolet, X-ray, and $\gamma$-ray bands.}
    \label{fig:fvar}
\end{figure}

The PDFs were fitted to the flux values of each light curve by maximum likelihood estimation (MLE). For the best-fit model selection, we employed the corrected Akaike Information Criterion (AICc; \citealt{Akaike7401, Sugiura1978, Liddle0705}) and the Bayesian Information Criterion (BIC; \citealt{Schwarz7807, Liddle0705}). The results of the fits are given in Table \ref{tab:fluxdist} and Figure \ref{fig:fluxdist}. The two models were compared using $\Delta{\rm AICc} = \Delta{\rm BIC}$, the difference between the value of a model and the lowest value. Following the criteria presented in \citet{Burnham2002}, models with $\Delta < 2$ were treated as competitive with the preferred one, while those with $\Delta > 10$ can be ruled out \citep{Psaradaki2010}. The lognormal model has the lower AICc and BIC in all light curves, with $\Delta{\rm AICc} = \Delta{\rm BIC}$ between $46.1$ (Swift-XRT) and $697.6$ (ZTF $r$) for the normal model. The normal distribution can therefore be ruled out in every band, and the lognormal is the preferred description of the flux distributions.  

The lognormal flux distributions have been reported for blazars in the X-ray \citep{Giebels2009Sep, Hervet2019May, Wani2023Jul}, VHE $\gamma$-ray \citep{HESSCollaboration2010Sep, Acciari2021Jun}, $\gamma$-ray \citep{Bhatta2020Mar, Kushwaha2017Nov, Shah2018Oct} and optical \citep{Bhatta2021Dec} bands. The lognormal flux distributions can originate in the accretion disc, in the fluctuating-accretion model where accretion rate variations produced at different radii modulate each other as they propagate inwards \citep{Lyubarskii1997Dec, Arevalo2006Apr}, and in blazars, lognormal distributions have been interpreted as a possible imprint of such disc fluctuations on the jet emission \citep{Bhatta2021Dec, Shah2018Oct}. Another scenario is within the jet,  where linear Gaussian fluctuations of the intrinsic particle acceleration or escape timescale can produce non-Gaussian fluctuations, including lognormal \citep{Sinha2018Oct}. Such a distribution may also arise when the observed flux is the sum of many compact regions that move in random directions within the jet, as in the minijets-in-a-jet model \citep{Giannios2009May, Biteau2012Dec}. The flux of a single region then follows a Pareto distribution, whereas the summed flux of many regions has a skewed distribution that is close to lognormal \citep{Biteau2012Dec}. The lognormal can also have a geometric, when changes in the viewing angle of the emitting region change the Doppler beaming \citep{Raiteri2024Dec}. While lognormal distribution may naturally arise from multiplicative variations,  additive variations have also been shown to produce such a distribution \citep[see][]{Scargle2020Jun}.

\subsection{Fractional Variability} \label{sec:fv}
The fractional variability amplitude ($F_{\rm var}$), was used to compare the variability between bands. It measures the intrinsic rms variability of a light curve relative to its mean flux, after subtracting the contribution of the measurement uncertainties, and being dimensionless, it can be compared between bands with different sensitivities \citep{Vaughan2003}. The $F_{\rm var}$ and its uncertainty are calculated following \citet{Bhatta2018} and listed in Table \ref{tab:fvar} and plotted in Figure \ref{fig:fvar}. The $F_{\rm var}$ follows a double-humped shape, similar to the shape of the broadband SED. The $\gamma$-ray band shows highest variability with $F_{\rm var} = 106.53\% \pm 1.07\%$, the optical and ultraviolet bands have $F_{\rm var} \sim 73\%-98\%$, and the X-ray band has the lowest variability with $F_{\rm var} = 64.50\% \pm 0.75\%$. Within the optical and ultraviolet bands, $F_{\rm var}$ increases towards shorter wavelengths, from the ZTF $i$ band to the $g$ band and from the UVOT $V$ band to the $W2$ band, where the small decrease in the $M2$ band is consistent with this trend within the uncertainties. A similar behaviour is seen in the X-ray range, where the hard band has a higher $F_{\rm var}$ than the soft band. Together with the Fermi-LAT value, the X-ray bands show an increase of $F_{\rm var}$ with energy. 

\begin{table}
\caption{Fractional variability amplitudes ($F_{\rm var}$), mean fluxes, and their associated uncertainties for the multi-wavelength light curves of {\src}.\label{tab:fvar}}
\centering
\let\tabular\savetabular
\begin{tabular}{@{}lcc@{}}
\toprule
Band & $F_{\rm var}$ (\%) & $\bar{F}$ \\
\midrule
ASAS-SN $V$ & $88.99 \pm 0.40$ & $(1.164 \pm 0.005) \times 10^{-11}$ \\
ASAS-SN $g$ & $76.98 \pm 0.27$ & $(1.887 \pm 0.005) \times 10^{-11}$ \\
ZTF $i$ & $72.97 \pm 0.23$ & $(2.638 \pm 0.006) \times 10^{-11}$ \\
ZTF $r$ & $80.31 \pm 0.13$ & $(1.959 \pm 0.003) \times 10^{-11}$ \\
ZTF $g$ & $83.30 \pm 0.20$ & $(1.595 \pm 0.003) \times 10^{-11}$ \\
Swift-UVOT $V$ & $87.84 \pm 0.71$ & $(2.321 \pm 0.016) \times 10^{-11}$ \\
Swift-UVOT $B$ & $89.91 \pm 0.79$ & $(1.868 \pm 0.015) \times 10^{-11}$ \\
Swift-UVOT $U$ & $90.19 \pm 0.88$ & $(1.649 \pm 0.015) \times 10^{-11}$ \\
Swift-UVOT $W1$ & $97.90 \pm 1.15$ & $(1.259 \pm 0.015) \times 10^{-11}$ \\
Swift-UVOT $M2$ & $96.60 \pm 1.71$ & $(1.087 \pm 0.019) \times 10^{-11}$ \\
Swift-UVOT $W2$ & $98.22 \pm 1.52$ & $(1.128 \pm 0.017) \times 10^{-11}$ \\
Swift-XRT Soft & $65.01 \pm 0.93$ & $(1.527 \pm 0.014) \times 10^{-1}$ \\
Swift-XRT Hard & $68.97 \pm 1.28$ & $(8.153 \pm 0.104) \times 10^{-2}$ \\
Swift-XRT Full & $64.50 \pm 0.75$ & $(2.359 \pm 0.018) \times 10^{-1}$ \\
Fermi-LAT & $106.53 \pm 1.07$ & $(1.264 \pm 0.013) \times 10^{-7}$ \\
\bottomrule
\end{tabular}
\par\vspace{4pt}
\noindent\parbox{\linewidth}{\textsc{Note:} $\bar{F}$ is given in the same units as the fluxes in Figure~\ref{fig:mwl_lc}.
}
\end{table}

Double-humped distributions of $F_{\rm var}$ have been reported for blazars of different classes \citep{Balokovic2016Mar, Rani2017Jan}. In HSP blazars, the X-rays show high $F_{\rm var}$, and the X-rays are synchrotron emission from the most energetic electrons \citep{Zhang2005Aug, Noel2022Sep}. HSPs also show a direct correlation between amplitude and energy \citep{MAGICCollaboration2020May, Ahnen2017Jul}. In LSP and ISP blazars whose soft X-ray band may contains the high-energy end of the synchrotron component, the soft band varies more than the hard band \citep{Foschini2006Sep}. For our source, the soft band does not vary more than the hard band, which indicates that the synchrotron component contributes little or none to the variable soft X-ray emission, in agreement with the hard X-ray spectra of the source.

\subsection{X-ray Spectral Analysis} \label{sec:xrayobs}
A total of 109 Swift-XRT observations were available for the source. We performed spectral fitting in 0.3-10 keV using three absorbed spectral models: power-law, log-parabola and broken power-law. For the best-fit model selection, we employed AICc and BIC, both of which weigh the improvement in the fit statistic against the model complexity. The AICc includes a finite-sample correction that is necessary here, given the small number of spectral bins in some of the observations \citep{Liddle0705}. The best-fitting parameters of the models, together with the preferred model by each criterion, are listed in Table \ref{tab:xrt_spectra}. Both criteria indicate that the simpler power-law is generally sufficient, with the log-parabola remaining competitive.

\begin{table*}
\caption{Best-fitting parameters of the spectral models for the Swift-XRT observations of {\src}, ordered by observation date.\label{tab:xrt_spectra}}
\centering
\let\tabular\savetabular
\setlength{\tabcolsep}{4.5pt}
\begin{tabular}{@{}lccccccccc@{}}
\toprule
 & \multicolumn{2}{c}{Power law (PL)} & \multicolumn{2}{c}{Log parabola (LP)} & \multicolumn{3}{c}{Broken power law (BPL)} & & \\
\cmidrule(lr){2-3} \cmidrule(lr){4-5} \cmidrule(lr){6-8}
ObsID & $\Gamma$ & $F_{\rm X\text{-}ray}$ & $\alpha$ & $\beta$ & $\Gamma_1$ & $E_{\rm b}$ & $\Gamma_2$ & TS$_{\rm curve}$ & AICc $|$ BIC \\
\midrule
00035381001 & $1.63\pm0.09$ & $4.32\pm0.37$ & $1.43\pm0.15$ & $0.47\pm0.27$ & $0.54\pm0.74$ & $0.72^{\dagger}$ & $1.75\pm0.24$ & 8.32 & LP $|$ LP \\
00036326001 & $1.78\pm0.16$ & $3.16\pm0.50$ & $1.61\pm0.24$ & $0.44\pm0.47$ & $-0.64\pm0.79$ & $1.23^{\dagger}$ & $2.00\pm0.31$ & 2.45 & LP $|$ PL \\
00036326002 & $1.81\pm0.15$ & $3.08\pm0.40$ & $1.73\pm0.22$ & $0.24\pm0.46$ & $1.34\pm6.43$ & $6.87^{\dagger}$ & $-3.00^{\dagger}$ & 0.76 & PL $|$ PL \\
\multicolumn{10}{c}{$\cdots$} \\
00019580001 & $1.17\pm0.21$ & $21.57\pm4.20$ & $0.95\pm0.43$ & $0.41\pm0.68$ & $-1.51\pm0.69$ & $0.21^{\dagger}$ & $1.17\pm0.21$ & 1.07 & PL $|$ PL \\
00019580002 & $1.29\pm0.22$ & $8.63\pm1.91$ & $1.33\pm0.39$ & $-0.08\pm0.64$ & $-3.00\pm1.50$ & $0.06^{\dagger}$ & $1.29^{\dagger}$ & 0.04 & PL $|$ PL \\
\bottomrule
\end{tabular}
\par\vspace{4pt}
\noindent\parbox{\linewidth}{\textsc{Note:} All models include Galactic absorption fixed at $N_{\rm H}=4.68\times10^{20}$\,cm$^{-2}$ and were fitted with \textit{cstat}. $F_{\rm X\text{-}ray}$ is the 0.3--10\,keV power-law flux in units of $10^{-12}$\,erg\,cm$^{-2}$\,s$^{-1}$, and $E_{\rm b}$ is the break energy in keV. Parameters marked $^{\dagger}$ have uncertainties that were not constrained by the fit. TS$_{\rm curve}$ is the curvature test statistic of the LP relative to the PL, and the last column gives the model preferred by the AICc and the BIC, respectively. The full table is available as supplementary material.}
\end{table*}

\begin{figure}
    \centering
    \includegraphics[width=1\linewidth]{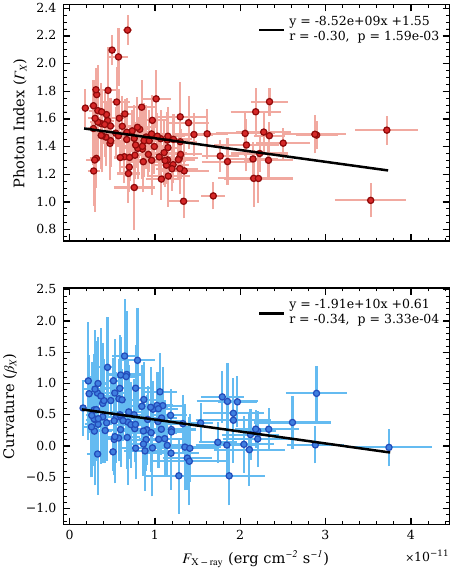}
    \caption{X-ray spectral parameters as a function of the 0.3--10 keV flux for the individual \textit{Swift}-XRT observations of {\src}. \textit{Top:} photon index $\Gamma_{\rm X}$ from the power-law model against the flux. \textit{Bottom:} curvature parameter $\beta_{\rm X}$ from the log-parabola model against the flux. The black lines are linear least-squares fits.}
    \label{fig:xrt_corr}
\end{figure}

The dependence of the X-ray spectral parameters on the flux is shown in Figure \ref{fig:xrt_corr}. The photon index decreases with increasing flux, with a Pearson coefficient of $r=-0.30$ ($p=1.6\times10^{-3}$), which indicates a weak but significant harder-when-brighter trend. The curvature also decreases with increasing flux ($r=-0.34$, $p=3.3\times10^{-4}$), so the brighter spectra are closer to a power-law. The flux dependence of the X-ray spectral shape is expected to differ between blazars according to the emission component that dominates the band \citep{Giommi2021Nov}. For BL Lacs, the spectrum becomes harder-when-brighter in HBLs, with a curvature that shows no consistent correlation with the flux \citep{Kapanadze2017Aug, Khatoon2022Sep, Das2025Jun, Taylor2026Feb, Kizhakkekalam2026Jul}, and softer-when-brighter in IBLs and LBLs, where the synchrotron tail enters the X-ray band \citep{Giommi2021Nov, Navaneeth2025Dec}. In FSRQs, the band samples the low-energy part of the IC component, which produces hard spectra, and a harder-when-brighter trend is commonly found \citep{Gu2011Jun, Hayashida2012Aug, Sahakyan2021Jul}, whereas softer spectra appear when the synchrotron tail contributes at the lowest energies \citep{2026arXiv261121518K}. The harder-when-brighter trend, together with the decrease of the curvature towards a power-law shape in bright states, follows the behaviour expected for an IC-dominated X-ray band. In this case, the hard IC component dominates increasingly as the source brightens, while the soft component at the lowest energies becomes relatively weaker \citep{Abdo2010Jun}. 

\begin{table*}
\caption{X-ray spectral-fitting results for {\src} during flaring epochs, obtained using power-law (PL), log-parabola (LP) and broken power-law (BPL) models.
\label{tab:flare_xray_spec}}
\centering
\let\tabular\savetabular
\begin{tabular}{lccccc}
\toprule
Parameter & F1 & F2 & F3 & F4 & F5 \\
\midrule
$\Gamma$ & $1.43\pm0.03$ & $1.24\pm0.07$ & $1.34\pm0.06$ & $1.50\pm0.05$ & $1.52\pm0.12$ \\
\midrule
$\alpha$ & $1.31\pm0.05$ & $1.13\pm0.13$ & $1.01\pm0.12$ & $1.35\pm0.08$ & $1.39\pm0.20$ \\
$\beta$ & $0.24\pm0.09$ & $0.21\pm0.20$ & $0.65\pm0.20$ & $0.34\pm0.15$ & $0.30\pm0.35$ \\
\midrule
$\Gamma_1$ & $-1.92\pm0.04$ & $-3.00\pm1.50$ & $-3.00\pm1.50$ & $1.35\pm0.09$ & $-0.66\pm3.01$ \\
$\Gamma_2$ & $1.47\pm0.03$ & $1.24\pm0.04$ & $1.34\pm0.04$ & $1.84\pm0.18$ & $1.63\pm1.62$ \\
$E_{\rm b}$ & $0.50\pm0.05$ & $0.08\pm0.45$ & $0.13\pm0.47$ & $2.32\pm0.48$ & $1.09\pm0.00$ \\
\midrule
$F_{\rm X\text{-}ray}$ & $1.96\pm0.07$ & $0.93\pm0.08$ & $0.97\pm0.08$ & $1.09\pm0.06$ & $1.12\pm0.14$ \\
TS$_{\rm curve}$ & 20.09 & 3.12 & 33.14 & 15.17 & 2.04 \\
AICc $|$ BIC & BPL $|$ BPL & LP $|$ PL & LP $|$ LP & LP $|$ LP & PL $|$ PL \\
\bottomrule
\end{tabular}
\par\vspace{4pt}
\noindent\parbox{\linewidth}{\textsc{Note:} $\Gamma$: PL photon index; $\alpha$, $\beta$: LP photon index at the reference energy and curvature parameter; $\Gamma_1$, $\Gamma_2$: BPL photon indices below and above the break energy $E_{\rm b}$; $F_{\rm X-ray}$: Flux from the LP model; TS$_{\rm curve}$: curvature test statistic of the LP relative to the PL. Energies are in keV and fluxes in $10^{-12}$\,erg\,cm$^{-2}$\,s$^{-1}$.}
\end{table*}

The same models were fitted to the spectra of the flares obtained by combining all observations falling within each, and the results are listed in Table~\ref{tab:flare_xray_spec}. A curved model is preferred in four of the five flares under the AICc and in three under the BIC. The LP is selected for F3 and F4 under both criteria and for F2 under the AICc only, whereas in F1 the BPL is preferred under both criteria and F5 remains best described by a PL. Following \citet{Nolan1204}, we quantified the significance of the spectral curvature with the ${\rm TS}_{\rm curve} = C_{\rm PL} - C_{\rm LP}$, where $C_{\rm PL}$ and $C_{\rm LP}$ are the Cash statistics of the power-law and log-parabola fits and TS$_{\rm curve}>16$ is adopted as the threshold for a significant curvature. The flare spectra show a stronger curvature than the individual observations, with TS$_{\rm curve}$ exceeding the threshold in F1 and F3 and lying marginally below it in F4, whereas F2 and F5 remain consistent with a power-law. 

\subsection{$\gamma$-ray Spectral Analysis} \label{sec:gammarayspec}
\begin{figure*}
    \centering
    \includegraphics[width=0.8\linewidth]{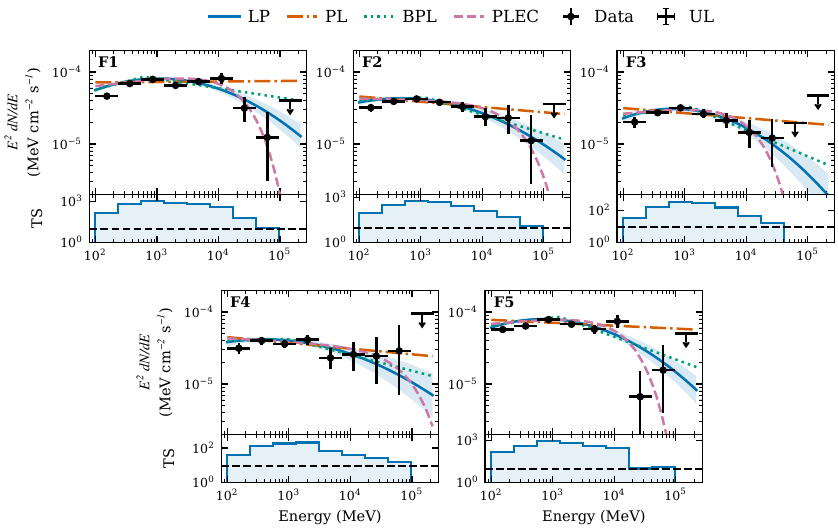}
    \caption{Fermi-LAT SED of {\src} in flaring epochs. Top: Flux points and upper limits (TS $\geq$ 9) with models: LP (solid blue), PL (dash-dotted orange), BPL (dotted green), PLEC (dashed pink); shaded band is the 1$\sigma$ LP uncertainty. Bottom: TS per bin from the LP fit; dashed line marks TS = 9.}
    \label{fig:flare_gamma_spec}
\end{figure*}

The $\gamma$-ray spectrum of each flaring epoch was fitted with four spectral models: power-law, log-parabola, broken power-law, and power-law with exponential cutoff. Spectral curvature was quantified with the \texttt{curvature()} method of \texttt{fermipy} as $\mathrm{TS}_{\rm curve}=2\left(\ln\mathcal{L}_{\rm LP}-\ln\mathcal{L}_{\rm PL}\right)$, where $\mathcal{L}$ is the likelihood of the fit with the LP or PL model \citep{Nolan1204}. The best-fit parameters of all models are listed in Table \ref{tab:flare_gamma_spec}, and the corresponding spectra with the models are shown in Figure \ref{fig:flare_gamma_spec}.

\begin{table*}
\caption{$\gamma$-ray spectral-fitting results for {\src} during flaring epochs, obtained from power-law (PL), log-parabola (LP), broken power-law (BPL), and power-law with exponential cutoff (PLEC) models.\label{tab:flare_gamma_spec}}
\centering
\let\tabular\savetabular
\begin{tabular}{lccccc}
\toprule
Parameter & F1 & F2 & F3 & F4 & F5 \\
\midrule
$\Gamma$ & $1.99\pm0.02$ & $2.07\pm0.03$ & $2.07\pm0.04$ & $2.08\pm0.05$ & $2.04\pm0.03$ \\
\midrule
$\alpha$ & $1.94\pm0.03$ & $2.04\pm0.03$ & $2.01\pm0.05$ & $2.05\pm0.06$ & $2.00\pm0.03$ \\
$\beta$ & $0.07\pm0.02$ & $0.05\pm0.02$ & $0.08\pm0.03$ & $0.05\pm0.03$ & $0.07\pm0.02$ \\
\midrule
$\Gamma_1$ & $1.77\pm0.09$ & $1.96\pm0.11$ & $1.89\pm0.11$ & $1.97\pm0.11$ & $1.89\pm0.06$ \\
$\Gamma_2$ & $2.13\pm0.06$ & $2.26\pm0.16$ & $2.36\pm0.17$ & $2.22\pm0.17$ & $2.31\pm0.10$ \\
$E_{\rm b}$ & $0.60\pm0.24$ & $1.19\pm1.48$ & $1.24\pm0.77$ & $1.03\pm1.20$ & $1.29\pm0.53$ \\
\midrule
$\Gamma_{\rm PLEC}$ & $1.89\pm0.03$ & $2.01\pm0.04$ & $1.92\pm0.07$ & $2.05\pm0.06$ & $1.92\pm0.04$ \\
$E_{\rm cut}$ & $23.79\pm7.13$ & $41.11\pm24.02$ & $12.73\pm6.06$ & $88.48\pm124.83$ & $17.32\pm5.57$ \\
\midrule
$F_{\gamma}$ & $7.07\pm0.46$ & $3.63\pm0.30$ & $2.35\pm0.24$ & $3.56\pm0.49$ & $6.51\pm0.44$ \\
TS$_{\rm curve}$ & 22.30 & 7.27 & 7.69 & 2.07 & 17.22 \\
\bottomrule
\end{tabular}
\par\vspace{4pt}
\noindent\parbox{\linewidth}{\textsc{Note:} $\Gamma$: PL photon index; $\alpha$, $\beta$: LP photon index at the reference energy and curvature parameter; $\Gamma_1$, $\Gamma_2$: BPL photon indices below and above the break energy $E_{\rm b}$; $\Gamma_{\rm PLEC}$, $E_{\rm cut}$: PLEC photon index and cutoff energy; $F_{\gamma}$: Flux from the LP model; TS$_{\rm curve}$: curvature test statistic of the LP relative to the PL. Energies are in GeV and fluxes in $10^{-10}$\,erg\,cm$^{-2}$\,s$^{-1}$.}
\end{table*}

The power-law photon index remains close to $\Gamma \simeq 2$ at every epoch, ranging from $1.99\pm0.02$ at F1 to $2.08\pm0.05$ at F4 showing hard spectrum. Significant curvature is found only at F1 and F5, the two epochs of highest flux, with $TS_{\mathrm{curve}} = 22.30$ and $17.22$, while others show below $3\sigma$. The curvature parameter, in contrast, does not change significantly between the flares with $\beta$ lies between $0.05$ and $0.08$. The broken power-law gives a hard index below the break, $\Gamma_1 = 1.77-1.97$, and a softer index above it, $\Gamma_2 = 2.13-2.36$, but the break energy is poorly constrained except at F1 and F5, where it lies between 0.6 and 1.3 GeV. The PLEC fits give photon indices that are harder than the corresponding power-law values and the cutoff energy is better constrained in F1 and F5 at $\sim20$ GeV. 

The photon index of $\Gamma \simeq 2$ found in all five flares is harder than the long-term average of $\Gamma = 2.19 \pm 0.01$ in the 4FGL catalogue \citep{Ajello2020Apr}. Hard LAT spectra were also reported during the flares of 2015 and 2025, which coincide with F1 and F5 \citep{Tanaka2016Aug, MAGICCollaboration2018Sep, Ouyang2026Nov}. These results indicate that the hardening of the $\gamma$-ray spectrum is a recurring feature of the high states of the source. The spectral parameters agree within their uncertainties in the flares separated by up to a decade, which suggests that the shape of the distribution of the electrons that produce the $\gamma$-ray emission does not change between the flares.

\subsection{Broad-band SED of Flares} \label{sec:sed}
The broadband SED of the flares are modeled with the standard one-zone leptonic emission scenario. The emission is assumed to originate from a compact, homogeneous, spherical plasmoid (blob) of radius $R$, located at a distance $R_H$ from the central supermassive black hole. The plasma within the blob is threaded by a uniform magnetic field of strength $B$ and moves downstream with a bulk Lorentz factor $\Gamma = (1 - \beta^2)^{-1/2}$, where $\beta = v/c$. As the blob moves relativistically at a small angle $\theta$ to the observer's line of sight, the emitted radiation is Doppler boosted, with the Doppler factor given by $\delta_D = 1/\Gamma (1 - \beta \cos \theta)$. The electron energy distribution (EED) is described by a broken power-law distribution,
\begin{equation}
N(\gamma) =
\begin{cases}
N_0 \, \gamma^{-p_1}, & \gamma_{\min} \leq \gamma \leq \gamma_{\mathrm{b}}, \\[6pt]
N_0 \, \gamma_{\mathrm{b}}^{\,p_2-p_1} \, \gamma^{-p_2}, & \gamma_{\mathrm{b}} < \gamma \leq \gamma_{\max},
\end{cases}
\label{eq:eed}
\end{equation}
where $N_0$ is the normalization of the electron number density in $\mathrm{cm^{-3}}$, $p_1$ and $p_2$ are the spectral indices below and above the break Lorentz factor $\gamma_{\mathrm{b}}$, respectively.

\begin{figure*}
    \centering
    \includegraphics[width=1.0\linewidth]{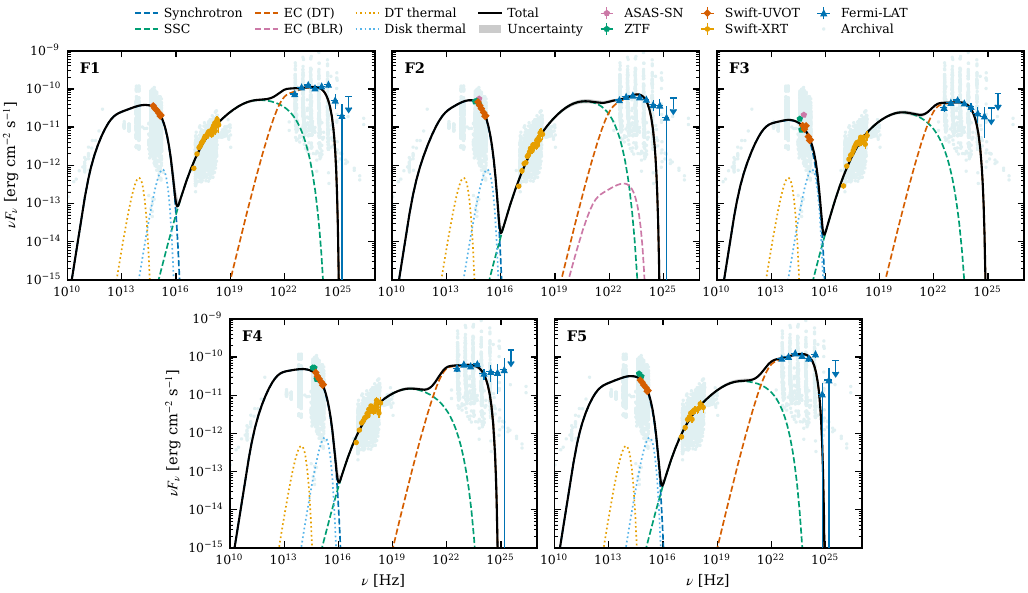}
    \caption{Broadband SED of {\src} in flaring epochs, modelled with a one-zone SSC+EC model. Dashed and dotted lines show the individual emission components and the solid black line is the total model, with the uncertainty in gray.}
    \label{fig:flare_sed}
\end{figure*}

\begin{table*}
\caption{Broadband SED modelling results for {\src} in flaring epochs, obtained with the one-zone SSC+EC model.\label{tab:flare_sed}}
\centering
\let\tabular\savetabular
\begin{tabular}{llccccc}
\toprule
Parameter & Units & F1 & F2 & F3 & F4 & F5 \\
\midrule
$\gamma_{\rm min}$ & & 229 & 242 & 274 & 183 & 189 \\
$\gamma_{\rm max}$ & & $7.05\times10^{3}$ & $4.08\times10^{3}$ & $5.43\times10^{3}$ & $4.34\times10^{3}$ & $5.11\times10^{3}$ \\
$N_0$ & cm$^{-3}$ & 331 & 427 & 721 & 163 & 341 \\
$\gamma_{\rm b}$ & & $4.17\times10^{3}$ & $2.57\times10^{3}$ & $3.26\times10^{3}$ & $2.57\times10^{3}$ & $4.26\times10^{3}$ \\
$p_1$ & & 2.62 & 2.38 & 2.76 & 2.81 & 2.6 \\
$p_2$ & & 3.62 & 3.45 & 3.82 & 3.98 & 3.6 \\
$R$ & cm & $1.37\times10^{16}$ & $7.12\times10^{15}$ & $7.13\times10^{15}$ & $1.30\times10^{16}$ & $1.07\times10^{16}$ \\
$R_{\rm H}$ & cm & $4.43\times10^{18}$ & $3.41\times10^{17}$ & $1.98\times10^{18}$ & $1.44\times10^{18}$ & $4.09\times10^{18}$ \\
$B$ & G & 0.399 & 0.697 & 0.423 & 0.745 & 0.424 \\
$\Gamma$ & & 15.3 & 22.9 & 16.3 & 24.4 & 28.6 \\
\midrule
$\chi^2_{\rm red}$ & & 1.91 & 1.44 & 3.95 & 2.11 & 1.73 \\
\midrule
$U_{\rm e}$ & erg\,cm$^{-3}$ & 0.141 & 0.202 & 0.334 & 0.0502 & 0.121 \\
$U_B$ & erg\,cm$^{-3}$ & $6.34\times10^{-3}$ & $1.93\times10^{-2}$ & $7.13\times10^{-3}$ & $2.21\times10^{-2}$ & $7.16\times10^{-3}$ \\
$U_{\rm e}/U_B$ & & 22.3 & 10.5 & 46.8 & 2.27 & 17 \\
$P_{\rm jet}$ & erg\,s$^{-1}$ & $2.76\times10^{45}$ & $2.26\times10^{45}$ & $1.85\times10^{45}$ & $3.13\times10^{45}$ & $5.90\times10^{45}$ \\
$\nu_{\rm peak}^{\rm syn}$ & Hz & $2.09\times10^{14}$ & $2.09\times10^{14}$ & $9.64\times10^{13}$ & $9.64\times10^{13}$ & $1.87\times10^{14}$ \\
$\nu_{\rm peak}^{\rm IC}$ & Hz & $4.48\times10^{23}$ & $3.59\times10^{23}$ & $2.06\times10^{23}$ & $1.48\times10^{23}$ & $4.01\times10^{23}$ \\
CD & & 2.97 & 1.42 & 2.98 & 1.30 & 3.82 \\
\bottomrule
\end{tabular}
\par\vspace{4pt}
\noindent\parbox{\linewidth}{\textsc{Note:} $\gamma_{\rm min}$, $\gamma_{\rm b}$, $\gamma_{\rm max}$: minimum, break and maximum electron Lorentz factors; $N_0$: electron density; $p_1$, $p_2$: indices below and above $\gamma_{\rm b}$; $R$, $R_{\rm H}$: radius and distance from the black hole of the emission region; $B$: magnetic field; $\Gamma$: bulk Lorentz factor; $U_{\rm e}$, $U_B$: electron and magnetic energy densities; $P_{\rm jet}$: jet power; $\nu_{\rm peak}^{\rm syn}$, $\nu_{\rm peak}^{\rm IC}$: synchrotron and IC peak frequencies; CD: Compton dominance.}
\end{table*}

The SED modelling was performed with the open-source code \texttt{JetSeT} \citep{Tramacere2020Sep, Tramacere2011Oct, Tramacere2009Jul}. The low-energy emission, extending from radio to ultraviolet, is dominated by synchrotron radiation, while the high-energy emission is from IC scattering, which comprises SSC and EC components. In the SSC process, seed photons are synchrotron photons produced within the emitting region, whereas EC involves seed photons from the BLR and DT. The external photon fields are characterised by a disk luminosity of $L_{\mathrm{Disk}} = 5.3 \times 10^{44}\,\mathrm{erg\,s^{-1}}$ \citep{BecerraGonzalez2021Jul}. The BLR is modelled with a covering factor of $\tau_{\mathrm{BLR}} = 0.1$, while the DT has a covering factor of $\tau_{\mathrm{DT}} = 0.6$ and emits thermal radiation at a temperature of $T_{\mathrm{DT}} = 1500\,\mathrm{K}$. The model has ten free parameters, including the EED parameters $\gamma_{\rm min}$, $\gamma_{\rm b}$, $\gamma_{\rm max}$, $N_0$, $p_1$, and $p_2$, and the physical parameters $R$, $R_{\rm H}$, $B$, and $\Gamma$. The size of the emission region is constrained by the variability timescale of each flare obtained in Section~\ref{sec:tvar}. For each flare, the initial parameter values were obtained from phenomenological model constraining and fitting was performed using \texttt{LSB}, followed by \texttt{Minuit} optimisation. Systematic uncertainties were added to the flux points, and the Fermi-LAT upper limits were not included in the fit. The model uncertainties were then estimated by the Markov chain Monte Carlo method, starting from the \texttt{Minuit} best-fit.

The broadband SEDs of the flares, together with the best-fitting model and their individual radiative components, are shown in Figure \ref{fig:flare_sed}, and the corresponding parameters are listed in Table \ref{tab:flare_sed}. In all five flares, the high-energy component is produced by EC scattering of dusty torus photons (EC-DT) with a subdominant SSC contribution, and EC-BLR is negligible. Synchrotron emission accounts for the radio to ultraviolet band, with an observed peak at $\nu^{Syn}_{\rm peak}\sim(0.96-2.09)\times10^{14}$ which places the source at the boundary between the LSP and ISP regimes, set at $10^{14}$\,Hz, with F3 and F4 at the threshold and F1, F2 and F5 above it. All five flares are particle dominated, with $U_e/U_B$ between 2.3 (F4) and 46.8 (F3). The total jet power, computed assuming one cold proton per radiating electron, is $P_{\rm Jet}=(1.9-5.9)\times10^{45}$\,erg\,s$^{-1}$ and F5 is the most powerful flare, while F3 has the lowest. The relative strength of the two SED components is quantified by the Compton dominance, ${\rm CD} = {\nu F_{\nu}}_{\rm IC}^{\rm peak}/{\nu F_{\nu}}_{\rm syn}^{\rm peak}$. The two components have comparable peak luminosities in F2 and F4 (${\rm CD} = 1.4$ and 1.3), whereas the IC component dominates in F1, F3 and F5 (${\rm CD} = 3.0-3.8$). The emission regions are compact, $R=(0.71-1.4)\times10^{16}$ cm, and close to the causality limit in all flares. The upper limit $R_{\rm max}$ is calculated using $\delta_D=23.8-28.7$ from the fitted bulk Lorentz factors $\Gamma=15.3-28.6$ for the fixed viewing angle $\theta=2^\circ$, which gives $R_{\rm max}=(0.8-1.7)\times10^{16}$ cm. The magnetic field strength ranges from $B=0.40$ G (F1) to $0.75$ G (F4) ahd the distance from the black hole are $R_H=(0.34-4.4)\times10^{18}$ cm. The shape of the electron distribution is similar in all flares. The index below the break is $p_1=2.38$--$2.81$ and the index above it $p_2=3.45$--$3.98$, so that the distribution steepens by $\Delta p=p_2-p_1=1.00$--$1.17$ at $\gamma_{\rm break}=(2.57$--$4.26)\times10^{3}$, and it extends from $\gamma_{\rm min}=183$--$274$ to $\gamma_{\rm max}=(4.08$--$7.05)\times10^{3}$. The radius--luminosity scaling gives $R_{\rm BLR, in}\simeq7\times10^{16}$ cm and $R_{\rm DT}\simeq4.6\times10^{18}$ cm. The emission regions are therefore located beyond the BLR and close to $R_{\rm DT}$ in all flares, with the F2 region closest to the BLR at $\simeq 5\,R_{\rm BLR,in}$.

The ratio $\psi=R/R_H$ is the half-angle by the emission region and equals the intrinsic half-opening angle of the jet $\theta_{\rm j}$, if the region fills the jet cross-section. For the flares, we find $\psi\simeq0.15^{\circ}-1.2^{\circ}$. \citet{Jorstad2017Sep} reported $\theta_{\rm j}=3.0^{\circ}\pm0.8^{\circ}$ and $0.5^{\circ}\pm0.6^{\circ}$ for the source, from the kinematics of the jet at 43 GHz using two methods. The values of $\psi$ suggest that in all flares dissipation occurs in a region smaller than the jet cross-section. F1, F3 and F5, with $\psi\simeq0.15^{\circ}-0.21^{\circ}$, lie below both estimates. Although F2 ($\psi\simeq1.2^{\circ}$) and F4 ($\psi\simeq0.52^{\circ}$) are compatible with the smaller estimate of $\theta_{\rm j}$, their regions are unlikely to fill the jet, since this estimate is poorly constrained, with an uncertainty comparable to its value and the flares would require different opening angles of the jet. A similar result follows from the relation $\Gamma\theta_{\rm j}\simeq0.2$ found by \citet{Jorstad2017Sep} for $\gamma$-ray bright blazars. F1, F3 and F5 have $\Gamma\psi\simeq0.05-0.08$, well below this value, and F4 ($\Gamma\psi\simeq0.22$) lies close to it. F2 ($\Gamma\psi\simeq0.48$) exceeds it, which requires a half-opening angle larger than typical and supports the larger estimate of $\theta_{\rm j}$ for the source.

The Mg\,\textsc{ii} line of the source, observed in 2015, has an equivalent width below 5 \AA{} \citep{BecerraGonzalez2021Jul}, which classifies it as a BL Lac object. However, $L_{\rm BLR}\simeq5.3\times10^{43}$\,erg\,s$^{-1}$ and $M_{\rm BH}\simeq2.3\times10^{8}\,M_\odot$ \citep{BecerraGonzalez2021Jul}, corresponding to $L_{\rm Edd}\simeq2.9\times10^{46}$\,erg\,s$^{-1}$, give $L_{\rm BLR}/L_{\rm Edd}\simeq1.8\times10^{-3}$, above the dividing value of $\sim5\times10^{-4}$ proposed by \citet{Ghisellini2011Jul} for the FSRQ--BL Lac separation. For $L_{\rm BLR}\simeq0.1\,L_{\rm disc}$, this dividing value corresponds to $L_{\rm disc}\sim5\times10^{-3}\,L_{\rm Edd}$, approximately where the disc goes from a radiatively efficient to an inefficient regime \citep{Ghisellini2011Jul}. The disc of the source lies above this value ($L_{\rm disc}\sim1.8\times10^{-2}\,L_{\rm Edd}$), consistent with a radiatively efficient disc and with the external photon fields required by the SED modelling. Within the factor of 3.5 uncertainty on $M_{\rm BH}$, the ratio $L_{\rm BLR}/L_{\rm Edd}$ decreases to $\sim5\times10^{-4}$, at the dividing value, consistent with being a transitional blazar \citep{BecerraGonzalez2021Jul}. 

\section{Discussion} \label{sec:discussion}
The multi-wavelength analysis of the source shows a broadband nature characterised by long-term variability and outbursts across the electromagnetic spectrum. The energy dependence of $F_{\rm var}$ follows the double-humped shape of the SED. The optical and ultraviolet bands lie near the synchrotron peak and the $\gamma$-rays cover the IC peak, so that both are produced by electrons near and above $\gamma_{\rm b}$, whereas the XRT band, at the low-energy end of the SSC component, is produced by electrons around $\gamma_{\rm min}$. The variability amplitude of the source therefore increases with the energy of the radiating electrons \citep{Castignani2017May, Rani2017Jan}. The intra-night optical variability was recorded during outbursts within F1 \citep{Bachev2015Jul}, F2, and F3 \citep{Raiteri2023Dec} and the size of the optical emission region is reported as $R < 3 \times 10^{14}$ cm. The optical variability originates in structures smaller than the regions that produce the high-energy flares \citep{Raiteri2023Dec}. The optical and $\gamma$-ray light curves are correlated without a measurable delay in 2022 \citep{Raiteri2023Dec} and during 2020--2023 \citep{Vlasyuk2023Dec}, whereas during F5 in 2025 the $\gamma$-rays lead the optical emission by $\simeq3.2$ days \citep{Ouyang2026Nov}. These correlations indicates that the optical emission region is closely linked to the high-energy dissipation zone, consistent with a shock propagating along the jet \citep{Marscher1985, Ouyang2026Nov}. These results also constrain the origin of the lognormal flux distribution. Accretion-rate fluctuations in the disc \citep{Lyubarskii1997Dec, Arevalo2006Apr} may contribute to the long-term variations, but the variability on timescales of hours to minutes arises in the jet. Within the jet, fluctuations of the acceleration or escape timescale \citep{Sinha2018Oct} can account for both the lognormal form and the larger amplitude of the bands produced by the most energetic electrons. The long-term variations of the source have been attributed to changes in the Doppler factor \citep{Raiteri2021Jul, Raiteri2023Dec, Pandey2023Nov}. The particle dominance of all flares favours shock in the jet, as inferred from the optical polarisation \citep{Morozova2014Sep, Raiteri2023Dec, Kouch2025Mar} and from the intra-day optical variability \citep{Raiteri2021Jul, Pandey2023Nov} and from the association of flares with newly ejected superluminal knots \citep{MAGICCollaboration2018Sep, Ouyang2026Nov}, although magnetic reconnection scenarios have also been proposed for flux and polarization variability \citep{Raiteri2023Dec}. The lognormal flux distribution of the source may therefore arise from intrinsic fluctuations in a particle-dominated emission region, combined with changes in its Doppler factor.

The X-ray and $\gamma$-ray spectra probe the two ends of the same electron distribution. The X-ray spectral variability, produced by IC emission of the electrons near $\gamma_{\rm min}$, resembles that of FSRQs. In the $\gamma$-ray band, the photon index and curvature of the Fermi-LAT spectra remain nearly constant in all five flares. The electron distribution steepens by $\Delta p\simeq1$ at $\gamma_{\rm b}$ in all flares, as expected for a radiative cooling break \citep{Georganopoulos2001Nov}. These results indicate that the shape of the distribution is set by the balance between radiative cooling and escape of the electrons, and that the acceleration process did not change between the outbursts. F1 and F5, separated by a decade, have nearly the same electron distribution and are located at nearly the same distance from the black hole and close to $R_{\rm DT}$. This recurrence suggests a fixed emission site, such as a standing shock, that is crossed repeatedly by disturbances propagating down the jet \citep{Morozova2014Sep, MAGICCollaboration2018Sep}. The Compton excess of F1, F3 and F5 agrees with the EC scattering of torus photons required to model the outbursts of 2015 and 2025 \citep{Tanaka2016Aug, MAGICCollaboration2018Sep, Ouyang2026Nov}, which correspond to F1 and F5. In contrast, F2 and F4, in which the two components have comparable luminosities, resemble the states that SSC reproduces without an EC component, such as the 2023 state observed with IXPE during F4 \citep{Kouch2025Mar} and the states before and after the 2025 outburst \citep{Ouyang2026Nov}. The $\gamma$-ray variability on timescales of 3.5--7.8\,hr also favours an EC origin of the $\gamma$-rays, since a dominant SSC component requires longer variability timescales \citep{Hu2024Mar}. Emission above 100 GeV is not restricted to the intermediate and low states in which the four LAT photons arrived, but also occurs within the outbursts. On the day of the VHE detection within F1, the LAT spectrum hardened to $\Gamma\approx1.8$ \citep{MAGICCollaboration2018Sep} and the X-ray spectrum softened \citep{Tanaka2016Aug}, both of which indicate a temporary extension of the electron distribution to higher energies. Outside the outbursts, the photons above 100 GeV may arise from EC cooling in a separate region.

The energetics and the broadband properties place the source between the two blazar classes. Its central engine is that of an FSRQ, with $L_{\rm BLR}/L_{\rm Edd}$ above the dividing value between the two classes and jet powers comparable to the accretion power. The Mg\,\textsc{ii} equivalent width below 5\,\AA{} can be attributed to the dilution of the broad lines by the beamed continuum, as in the masquerading BL Lacs \citep{Padovani2019Mar, Padovani2022Apr} or Changing-Look blazars \citep{Ruan2014, Mishra2021, Pandey2501, Chen2512, Azzollini2511}, which are intrinsically FSRQs whose broad lines are swamped by the Doppler-boosted jet emission. The EC origin of the $\gamma$-rays in all five flares supports this interpretation, as external photons are also required to reproduce the SEDs of most LSP BL Lacs, including that of the source in 2010, and these objects have therefore been proposed to be masquerading BL Lacs \citep{Hu2024Mar}. The emission properties are intermediate in the same way, the synchrotron peak lies at the boundary between the LSP and ISP classes, the Compton dominance places the source between FSRQs and BL Lacs \citep{Finke2013Feb}, and the X-ray spectral variability is that of FSRQs, whereas the emission above 100\,GeV is rare among FSRQs \citep{Paliya2025Sep}. The source can therefore be described as a transitional blazar \citep{BecerraGonzalez2021Jul} with an FSRQ-like central engine and BL Lac-like optical spectrum.

\section{Conclusions} \label{sec:conclusion}
We have presented a multi-wavelength study of the blazar {\src} based on observations from 2008 to 2026, focused on five high-energy flares (F1--F5) between 2015 and 2025. All five flares show $\gamma$-ray variability with shortest flux doubling/halving timescales of 3.5--7.8 hr, and the SED models place the emission in compact regions of $\sim10^{16}$ cm. The flux from the optical to the $\gamma$-ray band follows a lognormal distribution, and the fractional variability increases with the energy of the radiating electrons, from a minimum in the X-ray band to a maximum in the optical-ultraviolet and $\gamma$-ray bands. We suggest that the lognormal distribution arises from intrinsic fluctuations in a particle-dominated emission region, combined with changes in its Doppler factor.

The flare SEDs are reproduced by a one-zone leptonic model in which the $\gamma$-rays are produced by EC scattering of dusty torus photons and the X-rays by SSC emission. The X-ray spectra show a weak but significant harder-when-brighter trend, and the LAT spectra remain hard ($\Gamma\simeq2$) and nearly unchanged in all flares. The electron distributions of all flares steepen by $\Delta p\simeq1$ at $\gamma_{\rm b}$, as expected for a cooling break. These results indicate that the acceleration process did not change between the outbursts. The recurrence of flares at nearly the same distance from the black hole suggests a standing shock that is crossed repeatedly by disturbances propagating down the jet. The detection of four LAT photons with energies of 123--188 GeV places their emitting region beyond the BLR. Their arrival is outside the flares, in intermediate and low states. With $L_{\rm BLR}/L_{\rm Edd}\simeq1.8\times10^{-3}$, above the dividing value between FSRQs and BL Lacs, the source has an FSRQ-like central engine, while its synchrotron peak and Compton dominance are intermediate between the two classes. Together with its BL Lac-like optical spectrum, these properties indicate that {\src} is a transitional blazar.

\begin{acknowledgments}
\textbf{\textit{Acknowledgments}}: This work was partially supported by a program of the Polish Ministry of Science under the title “Regional Excellence Initiative”, project No. RID/SP/0050/2024/1.\\
\textbf{\textit{Facilities}}: Fermi (LAT), Swift (XRT and UVOT), ASAS-SN, ZTF \\
\textbf{\textit{Software}}: Astropy, Fermipy, HEASoft, JetSeT. 
\end{acknowledgments}

%





\bibliography{apj}{}
\bibliographystyle{aasjournal}



\end{document}